\documentclass{article}

\usepackage{microtype}
\usepackage{graphicx}
\usepackage{natbib}
\usepackage{subcaption}
\usepackage{booktabs} 
\usepackage{listings}
\usepackage{listings}
\usepackage[hidelinks,colorlinks=true,linkcolor=blue,citecolor=blue,breaklinks=true]{hyperref}    
\usepackage[ruled,vlined]{algorithm2e} 
\usepackage{enumitem}
\usepackage{amsmath}

\usepackage[eandd, final]{neurips_2026}
 
\usepackage{amsmath}
\usepackage{amssymb}
\usepackage{mathtools}
\usepackage{amsthm}
\usepackage{color}

\usepackage[capitalize,noabbrev]{cleveref}

\usepackage{xurl} 
\usepackage{listings}
\usepackage{xcolor}

\theoremstyle{plain}

\theoremstyle{definition}

\theoremstyle{remark}

\usepackage[textsize=tiny]{todonotes}

\title{\textsc{MathlibPR}: Pull Request Merge-Outcome Benchmark for Formal Mathematical Libraries}
\author{%
  Zixuan Xie\thanks{Equal contribution} \\
  University of Virginia\\
  \texttt{xie.zixuan@email.virginia.edu} \\
  \And
  Xinyu Liu$^*$ \\
  University of Virginia\\
  \texttt{xinyuliu@virginia.edu} \\
  \And
  Shangtong Zhang \\
  University of Virginia\\
  \texttt{shangtong@virginia.edu} \\
}

\begin{document}

\maketitle

\begin{abstract}
The ecosystem of Lean and Mathlib has become the de facto standard for large language model (LLM) assisted formal reasoning with remarkable successes in recent years.
Those successes, however, only consume Mathlib as an essential dependency but do not directly contribute to it.
In the meantime, 
the growth of Mathlib has recently been bottlenecked by the review process, which requires human reviewers to judge whether proposed pull requests (PRs) follow the Mathlib's conventions and are worth integrating as part of a shared mathematical infrastructure.
This leads to our central question: can LLMs help review Mathlib PRs?
To this end,
we introduce \textsc{MathlibPR}, a benchmark built from real Mathlib4 PR histories. 
We further propose a staged evaluation protocol and use it to evaluate both LLM models (e.g., DeepSeek, Qwen, Goedel, and Kimina) and LLM agents (e.g., Codex and Claude Code).
Surprisingly, both LLM models and LLM agents struggle to distinguish merge-ready PRs from build-passing PRs that were revised or never merged.
By turning Mathlib PR histories into a supervised signal, \textsc{MathlibPR} provides a step toward reviewer assistants and reward models that could help evaluate PRs and steer LLMs toward producing merge-ready Mathlib contributions.
\end{abstract}

\section{Introduction}
Mathlib \citep{mathlib2020} has become the de facto standard for large language model (LLM) assisted formal reasoning. 
Built on Lean4 \citep{moura2021lean}
, Mathlib is both a large formal library and an active open-source project, with reusable definitions, theorems, tactics, and APIs spanning diverse areas of mathematics.

Recent successes around this ecosystem of Mathlib and Lean4 span several levels. 
MiniF2F \citep{zheng2021minif2f} makes formal Olympiad-level problems a standard benchmark for LLM reasoning. 
AlphaProof \citep{hubert2025olympiad} 
demonstrates the power of LLM-based formal reasoning in competition mathematics. 
The Gauss agent \citep{mathinc_gauss_2025} and its strong Prime Number Theorem formalization \citep{mathinc_strongpnt_2025} show the scale at which LLM formalization projects are now being attempted.
Various areas of machine learning theory have also been formalized in Lean, including optimization \citep{li2024optimization}, reinforcement learning \citep{zhang2025rllean}, and statistical machine learning \citep{sonoda2025lean,zhang2026sltlean} 


These successes, however, only consume Mathlib as an essential dependency but do not directly and systematically contribute to it. 
This gap is precisely what surfaced in a recent Lean Zulip discussion around the LLM-assisted De Giorgi--Nash--Moser formalization \citep{lean_zulip_dgnm}. 
The discussion contrasts code that compiles and certifies a target theorem with code that is reusable, maintainable, stated at the right level of generality, and suitable for integration into Mathlib.
Although the correctness of LLM-generated code can be checked by the Lean kernel, whether such code is merge-ready for Mathlib is a more complex judgment that requires human review.
To our knowledge, the only LLM-based project that is specifically targeting the growth of Mathlib is MathlibLemma \citep{liu2026mathliblemma}, which studies the automatic discovery and formalization of missing folklore lemmas and reports that a subset of its verified outputs has already been merged into Mathlib.
However, \citet{liu2026mathliblemma} also acknowledge that verified proofs require further human review before becoming mergeable Mathlib contributions.


Unfortunately, human review is difficult to scale. 
The Mathlib Initiative roadmap reports that Mathlib now contains over 1.9 million lines of formally verified mathematics written by more than 500 contributors \citep{mathlib_roadmap}.
It also identifies the review queue as the primary constraint on Mathlib's growth, with roughly 300 pull requests (PRs) in backlog and median wait times around two weeks \citep{mathlib_roadmap}. 
Each PR requires human reviewers to judge whether the proposed code fits the surrounding library, follows Mathlib conventions, and is worth integrating as part of a shared mathematical infrastructure.
This leads to our central question: can LLMs help review Mathlib PRs?

To study this question, we introduce \textsc{MathlibPR}, a benchmark for evaluating whether LLMs can predict the merge outcomes of Mathlib PR snapshots.
Each example is drawn from a real Mathlib4 PR and has
already passed build checks.
\footnote{We use the public Mathlib4 repository,
\href{https://github.com/leanprover-community/mathlib4}{\texttt{leanprover-community/mathlib4}},
as the source of PR histories.}
\texttt{Merge-ready} examples correspond to snapshots accepted into Mathlib, while \texttt{not-merge-ready} examples correspond to build-passing snapshots that were later revised or never merged.
Since every example already builds, the benchmark rules out trivial strategies based on syntax errors or compilation failure and focuses on reviewer-like judgment.
Our contributions are threefold.
\begin{enumerate}[leftmargin=*,label=\arabic*.]
    \item We construct \textsc{MathlibPR} from real Mathlib4 PR histories, yielding a benchmark for predicting merge outcomes of build-passing Mathlib contributions.
    \item We propose a staged evaluation protocol that progressively supplies richer review-relevant context to evaluate LLMs on merge-readiness tasks.

    \item We conduct an empirical study of both LLM models (e.g., DeepSeek \citep{guo2025deepseek}, Qwen \citep{yang2025qwen3}, Goedel \citep{lin2025goedelv2}, and Kimina \citep{wang2025kimina}) and LLM agents (e.g., Codex \citep{openai2026codexcli} and Claude Code \citep{anthropic2026claudecode}) 
    , showing that reviewer-like judgment remains difficult for current systems even when every snapshot already passes build checks.
\end{enumerate}
More broadly, \textsc{MathlibPR} is intended as a step toward reviewer
assistants and reward models for formal mathematical libraries. 
Such models could help evaluate human-written PRs, triage LLM-generated contributions, and steer future LLM models and agents toward generating Lean code that is not only correct, but also maintainable, integrated, and merge-ready for Mathlib.

\section{Related Work}
\label{sec:related-work}

\paragraph{LLM-assisted formal theorem proving and benchmarks.}
Recent work has produced strong results in LLM-assisted formal 
theorem proving on Olympiad-level, undergraduate, and library-scale 
problems. Benchmarks such as MiniF2F~\citep{zheng2021minif2f}, 
ProofNet~\citep{azerbayev2023proofnet}, FIMO~\citep{liu2023fimo}, 
PutnamBench~\citep{tsoukalas2024putnambench}, and 
FormalMATH~\citep{yu2025formalmath} measure whether systems can 
generate proofs that are accepted by a proof assistant. 
LeanDojo~\citep{yang2023leandojo} extracts theorems and premises 
from Mathlib to support retrieval-augmented proving over a large 
formal library. Recent provers and agents, including 
DeepSeek-Prover~\citep{xin2024deepseekprover}, 
Goedel-Prover-V2~\citep{lin2025goedelv2}, 
Kimina-Prover~\citep{wang2025kimina}, 
Seed-Prover~\citep{chen2025seedprover}, and 
AlphaProof~\citep{hubert2025olympiad}, demonstrate that proof 
assistants provide a reliable correctness signal for evaluating 
and improving formal proof generation. \textsc{MathlibPR} differs 
in what it asks of a system. Once a Mathlib PR snapshot already 
builds, the question is no longer whether the kernel accepts the 
proof, but whether the contribution is suitable for integration 
into a maintained formal library.

\paragraph{Automated code review and pull-request evaluation.}
Automated code review has been studied in mainstream software 
engineering. CodeReviewer~\citep{li2022codereviewer} formulates 
review tasks such as code-change quality estimation, review 
comment generation, and code refinement, trained on large-scale 
open-source review data. More recent benchmarks such as 
SWRBench~\citep{zeng2025swrbench} and 
Sphinx~\citep{zhang2026sphinx} evaluate LLM-based PR review with 
project-level context and structured review criteria. Beyond 
review, SWE-bench~\citep{jimenez2024swebench} uses real GitHub 
issues and PRs to evaluate whether agents can edit a repository 
to resolve an issue. 
\textsc{MathlibPR} is closest to this line in spirit but differs in domain and signal. 
Its examples come from a formal mathematical library where every snapshot already passes build checks, and its task is to predict merge outcomes from code and immediate artifacts rather than rely on post-hoc review signals.

\paragraph{Library growth and contribution quality in Mathlib.}
Mathlib is both a corpus of verified theorems and a maintained 
open-source library whose contributions must remain reusable, 
documented, and coherent with surrounding 
APIs~\citep{mathlib2020}. Prior work on maintaining Mathlib 
emphasizes that kernel checking is only one part of library 
quality: linters, documentation, review practices, and 
project-level conventions are needed to reduce maintainer burden 
and preserve library coherence~\citep{vandoorn2020maintaining, 
baanen2025growing}. Several recent systems contribute to Mathlib 
growth from the generation side. 
MathlibLemma~\citep{liu2026mathliblemma} studies the automatic 
discovery and formalization of missing folklore lemmas, and 
reports that some generated lemmas have been upstreamed into 
Mathlib. Related systems synthesize useful lemmas or conjectures 
from proof traces, library contexts, or neuro-symbolic 
templates~\citep{sivaraman2022lemma, onda2025leanconjecturer, 
alhessi2025lemmanaid}, while other feedback-driven provers 
generate intermediate lemmas or subgoals as scaffolding for 
solving hard theorems~\citep{dong2024prodr, zhou2025deltaprover, 
ospanov2025apollo, varambally2025hilbert}. \textsc{MathlibPR} 
addresses the other side of library growth. Rather than 
generating new statements or proofs, it turns historical Mathlib 
review outcomes into a benchmark for evaluating whether 
build-passing contributions are merge-ready.

\section{Task Definition and Dataset Construction}
\label{sec:dataset}
\subsection{Merge-Outcome Prediction for Build-Passing PR Snapshots}
\label{sec:task-definition}

\textsc{MathlibPR} evaluates merge-outcome prediction for build-passing Mathlib4 PR snapshots.
A snapshot represents the state of a PR at a selected commit, together with the reconstructed code changes for that state. 
Given this snapshot-level context, the task is to decide whether the selected state is ready to be accepted into Mathlib4.
All selected snapshots pass the build checks by construction, and the benchmark deliberately withholds reviewer comments, discussion threads, and other post-hoc social signals that would otherwise leak the outcome. 
\textsc{MathlibPR} therefore probes the gap between build-passing code and merge-ready code, asking whether a system can make a reviewer-like judgment from the code and its immediate artifacts alone.

The dataset labels are binary.
A snapshot is labeled \texttt{merge\_ready} if it is the version accepted into Mathlib4, and \texttt{not\_merge\_ready} if it was either later revised before acceptance or belongs to a PR that was never accepted. 
The \texttt{not\_merge\_ready} label is snapshot-level: it indicates that the selected state was not merge-ready at that point in the PR history, not that the PR could never be merged.

\subsection{Dataset Construction Pipeline}
\label{sec:construction-pipeline}

\begin{figure}[t]
    \centering
    \includegraphics[width=\linewidth]{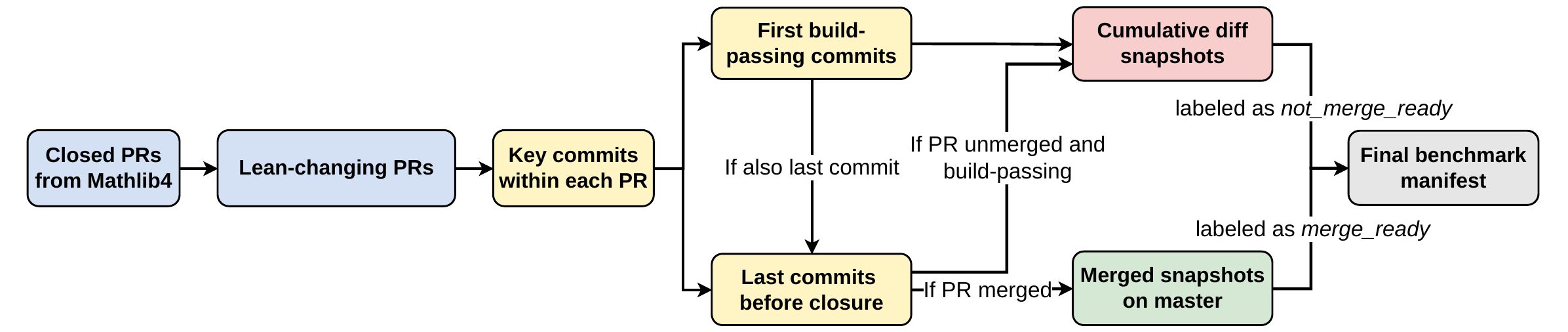}
    \caption{
    Overview of the \textsc{MathlibPR} dataset construction pipeline.
    Starting from closed PRs in the Mathlib4 repository, we retain PRs with Lean file changes.
    Historical build-check records identify build-passing commits on PR branches, including the first build-passing commit and, for unmerged PRs, the last build-passing commit before closure.
    These states provide \texttt{not\_merge\_ready} examples when they are later revised or never merged.
    For merged PRs, the \texttt{merge\_ready} example is the version accepted into Mathlib4 on the master branch.
    We also apply a temporal cutoff so that positive and negative examples come from a comparable period of Mathlib4 history.
    Cumulative diffs are reconstructed for the selected snapshots, and the resulting labeled examples form the final \textsc{MathlibPR} manifest.
    }
    \label{fig:dataset-pipeline}
\end{figure}

We construct \textsc{MathlibPR} from closed PRs in the Mathlib4 repository; 
Figure~\ref{fig:dataset-pipeline} summarizes the pipeline. 
We start from  closed PRs, retain those with Lean file changes, and use historical build-check records to recover build-passing states on PR branches. 
Because such records are not uniformly available throughout Mathlib4 history, they determine the period from which \texttt{not\_merge\_ready} examples can be reliably constructed.

For \texttt{not\_merge\_ready} examples, we draw two kinds of snapshots  from PRs with recovered build-check records: 
(i) the first build-passing  commit, when this state is later revised before acceptance; 
and (ii) the  last commit before closure of an unmerged PR, when that commit is build-passing. 
In both cases, we reconstruct the snapshot's cumulative  diff against the merge-base between the PR branch and its target Mathlib4 branch.

For \texttt{merge\_ready} examples, we take the version accepted into 
Mathlib4 on the master branch. 
To keep the two classes drawn from a  comparable period, we retain accepted snapshots only from the period beginning with the earliest retained \texttt{not\_merge\_ready} snapshot. 
When selected snapshots from the same PR coincide, we keep only one and assign its label by the rules above. 
The resulting labeled snapshots form  the final benchmark manifest, whose fields we describe in Appendix~\ref{app:access-manifest}.

\subsection{Dataset Statistics}
\label{sec:dataset-statistics}

We construct the current version of \textsc{MathlibPR} from closed PRs
available in the Mathlib4 repository as of April 5, 2026.
Table~\ref{tab:dataset-summary} summarizes the construction coverage and final
benchmark composition.
The final dataset contains 15,895 labeled build-passing snapshots from 12,063 PRs:
11,409 \texttt{merge\_ready} snapshots and 4,486 \texttt{not\_merge\_ready} snapshots.
Among merged PRs, the dataset includes 3,687 within-PR pairs, each consisting of an earlier build-passing snapshot and the version ultimately accepted into Mathlib4. 
These pairs enable a controlled within-PR analysis (Section~\ref{sec:within_pr_pairs}) that tests whether models can recognize revision progress on the same underlying contribution.
The benchmark is publicly available at \url{https://huggingface.co/datasets/MathlibPR/MathlibPR}, with a Croissant metadata file conforming to NeurIPS 2026 E\&D requirements.


\begin{table}[t]
\centering
\caption{Construction coverage and final benchmark composition for \textsc{MathlibPR}.}
\label{tab:dataset-summary}
\begin{tabular}{lrr}
\toprule
Component & Unit & Count \\
\midrule
\multicolumn{3}{l}{\textit{PR-level coverage}} \\
Closed Mathlib4 PRs as of April 5, 2026 & PRs & 33,444 \\
PRs with at least one Lean-file change & PRs & 24,172 \\
PRs represented in the final benchmark & PRs & 12,063 \\
\addlinespace
\multicolumn{3}{l}{\textit{Final benchmark composition}} \\
Labeled build-passing snapshots & Snapshots & 15,895 \\
\quad \texttt{merge\_ready} snapshots & Snapshots & 11,409 \\
\quad \texttt{not\_merge\_ready} snapshots & Snapshots & 4,486 \\
Within-PR pairs & Pairs & 3,687 \\
\bottomrule
\end{tabular}
\end{table}

\section{Experimental Setup}
\label{sec:experiments}

\subsection{Evaluated Systems}
\label{sec:evaluated-systems}

We evaluate two classes of systems on \textsc{MathlibPR}: LLM models, 
which receive only the supplied prompt, and LLM agents, which can 
additionally inspect a read-only local repository checkout.

\paragraph{LLM models.}
We first evaluate LLM models on the full benchmark. 
The evaluated models are the open-weight reasoning models DeepSeek-R1-Distill-Qwen (``DeepSeek'') and Qwen3-8B (``Qwen''); 
the library specialist Goedel-Prover-V2-32B (``Goedel''), a model pre-fine-tuned on Mathlib; 
and Kimina-Prover-Distill-8B (``Kimina''), a model trained via RL on formal competitions such as MiniF2F \citep{zheng2021minif2f}.

\paragraph{LLM agents.}
We additionally evaluate coding agents. Beyond the prompt, these agents 
can inspect a read-only local checkout at the snapshot commit and use 
coding tools to gather evidence before making their judgment. Because 
this setting is substantially more expensive, we evaluate
each agent at each stage on a balanced subset of 500 \texttt{merge\_ready} and 500 \texttt{not\_merge\_ready} 
examples rather than on the full benchmark.

We evaluate Codex via \texttt{codex exec} with GPT-5.4, and Claude Code 
via the local Claude CLI backed by Claude Sonnet 4.6. Each sample is 
processed in a fresh per-sample CLI session, so no state is carried 
across examples. Agents are restricted to read-only repository 
inspection: they cannot edit files, install dependencies, or access web 
resources, GitHub or PR pages, remote APIs, or git history. These 
restrictions are essential, because otherwise the task can be 
short-circuited by retrieving the known PR outcome instead of judging 
merge-readiness from the provided artifacts.

\subsection{Three-Stage Evaluation Protocol}
\label{sec:staged-protocol}

We evaluate LLM models and agents on \textsc{MathlibPR} under a three-stage protocol that supplies progressively richer evidence for each snapshot. 
All stages share the same review prompt contract and structured output schema; only the supplied evidence changes from Stage~1 to Stage~3. 
This design lets us ask not only whether current systems can make reviewer-like judgments, but also whether those judgments improve as richer review context is supplied.

Stage~1 provides code-local evidence for the snapshot: 
the diff itself, the changed-file context, and compact digests of relevant Mathlib review guidance on naming, style, and 
documentation.
\footnote{See 
\url{https://leanprover-community.github.io/contribute/naming.html}, 
\url{https://leanprover-community.github.io/contribute/style.html}, and 
\url{https://leanprover-community.github.io/contribute/doc.html}.} 
Stage~2 adds structured diagnostics covering linting, imports, declaration placement, documentation, and API fit. 
These diagnostics largely come from the same commit-level GitHub check records used to identify build-passing states in Section~\ref{sec:construction-pipeline}, 
repurposed as review signals. 
For accepted snapshots merged into master, comparable historical diagnostics are not always recoverable in the same form.
In those cases, we backfill linting diagnostics locally. 
Diagnostics are treated as evidence rather than as oracle labels: they may surface potentially review-relevant issues, but the system must still determine whether those issues actually block merge-readiness. 
Stage~3 further adds the PR title and description as stated intent.

The same staged restriction applies to LLM agents. Although they may inspect the local repository, our packaged diagnostics and PR intent are supplied only at Stages~2 and~3 respectively.
Full prompt templates and input field definitions are deferred to Appendix~\ref{app:prompts-schema}.

\subsection{Structured Output Schema}
\label{sec:outputs-validity}

All systems are required to return a structured JSON review rather than free-form commentary.
This follows prior LLM-evaluation work that uses explicit criteria, form-filling, or customized score rubrics to make judgments more structured, comparable, and aligned with human assessment \citep{liu2023geval,kim2024prometheus,hashemi2024llmrubric}.
The output includes an overall verdict, a scalar merge-readiness score, confidence fields, axis-level assessments along eight rubric axes, and brief supporting fields such as major strengths, blockers, and minimal required changes. 
The eight axes operationalize the concerns emphasized in Mathlib's reviewer and contributor documentation\footnote{See Mathlib's PR review guide 
(\url{https://leanprover-community.github.io/contribute/pr-review.html}) 
along with the naming, style, and documentation guidelines linked 
in Section~\ref{sec:staged-protocol}.} into a structured 
code-centric rubric. Outputs are validated against a fixed schema. 
The eight axes and full schema are detailed in 
Appendix~\ref{app:prompts-schema}.


An output that fails this validation is treated as invalid, meaning it cannot be parsed, violates the schema, omits required fields, or exceeds the per-sample runtime budget. 
Invalid outputs reflect interface or execution failures rather than reviewer decisions. 
A schema-valid output may still abstain at the overall-verdict level via \texttt{uncertain}, which reflects a substantive decision not to commit rather than a failure to produce a usable review.




\subsection{Metrics}
\label{sec:metrics}

A system's output is either invalid or carries a verdict of 
\texttt{merge\_ready}, \texttt{not\_merge\_ready}, or \texttt{uncertain}. 
We report four primary metrics on these outputs.

\paragraph{MR and NMR Recall.}
MR Recall is recall on \texttt{merge\_ready} examples, and NMR Recall is recall on \texttt{not\_merge\_ready} examples. 
We count a prediction as correct only if the verdict matches the reference label. 
Outputs with verdict \texttt{uncertain} and invalid outputs are both counted as errors. 
Reporting the two recalls separately exposes class-asymmetric behavior that a single accuracy metric would hide.

\paragraph{Balanced accuracy.}
We also report balanced accuracy as the mean of the two recalls,
\[
\mathrm{Bal.\ Acc.} = \frac{\mathrm{MR\ Recall} + \mathrm{NMR\ Recall}}{2}.
\]
A high MR Recall or NMR Recall on its own can mask poor performance on the other class, so balanced accuracy summarizes both in a single number.

\paragraph{Valid-output rate.}
Valid-output rate is the fraction of examples for which the system returns a schema-valid output within the per-sample runtime budget. 
This metric is informative because some systems fail by timing out or producing malformed outputs rather than by committing to an incorrect verdict.

\paragraph{AUROC.}
We also report the area under the receiver operating characteristic curve (AUROC), computed from the scalar merge-readiness scores in schema-valid outputs. 
AUROC measures how well the system ranks \texttt{merge\_ready} examples above \texttt{not\_merge\_ready} examples, independent of any discrete decision threshold. 
Because invalid outputs do not yield usable scores, AUROC should be read alongside the valid-output rate.

\section{Main Results}
\label{sec:main}
Table~\ref{tab:main_results_all} reports the main results for both LLM models and LLM agents. 
Because LLM models are evaluated on the full benchmark while LLM agents are evaluated on a balanced subset of 500 positive and 500 negative examples, numbers should not be compared directly across the two classes.
\begin{table*}[t]
\centering
\small
\setlength{\tabcolsep}{4.5pt}
\begin{tabular*}{\textwidth}{@{\extracolsep{\fill}}lllccccc}
\toprule
Model & Eval.\ set & Stage & AUROC & MR Recall & NMR Recall & Bal. Acc. & Valid rate \\
\midrule
\multicolumn{8}{l}{\textit{LLM models}} \\
DeepSeek & Full & S1 & 40.5 & 32.6 & 3.5 & 18.1 & 98.3 \\
DeepSeek & Full & S2 & 51.4 & 61.7 & 3.0 & 32.4 & 97.2 \\
DeepSeek & Full & S3 & 52.4 & 69.8 & 2.3 & 36.0 & 92.9 \\
Qwen & Full & S1 & 49.9 & 31.6 & 0.0 & 15.8 & 96.6 \\
Qwen & Full & S2 & 49.7 & 38.7 & 0.0 & 19.3 & 96.2 \\
Qwen & Full & S3 & 48.4 & 48.3 & 0.0 & 24.1 & 97.0 \\
Goedel & Full & S1 & 30.0 & 0.1 & 0.3 & 0.2 & 68.3 \\
Goedel & Full & S2 & 51.3 & 2.5 & 0.3 & 1.4 & 18.9 \\
Goedel & Full & S3 & 53.1 & 1.7 & 0.2 & 1.0 & 13.5 \\
Kimina & Full & S1 & 49.5 & 0.1 & 18.6 & 9.3 & 18.8 \\
Kimina & Full & S2 & 51.9 & 0.0 & 8.2 & 4.1 & 9.5 \\
Kimina & Full & S3 & 51.6 & 0.0 & 9.0 & 4.5 & 11.3 \\
\midrule
  \multicolumn{8}{l}{\textit{LLM agents}} \\
  Codex & 500/500 & S1 & 63.2 & 91.6 & 14.6 & 53.1 & 100.0 \\
  Codex & 500/500 & S2 & 58.9 & 91.4 & 9.4 & 50.4 & 100.0 \\
  Codex & 500/500 & S3 & 61.3 & 92.0 & 12.0 & 52.0 & 99.9 \\
  Claude Code & 500/500 & S1 & 68.8 & 29.8 & 14.0 & 21.9 & 48.8 \\
  Claude Code & 500/500 & S2 & 65.9 & 36.6 & 10.6 & 23.6 & 55.6 \\
  Claude Code & 500/500 & S3 & 66.2 & 38.0 & 12.0 & 25.0 & 56.2 \\
\bottomrule
\end{tabular*}
\caption{
Main results for all evaluated systems. All values are percentages, with AUROC multiplied by 100 and computed on schema-valid outputs with usable scores. 
MR Recall and NMR Recall are recall on \texttt{merge\_ready} and \texttt{not\_merge\_ready} examples, with \texttt{uncertain} and invalid outputs counted as errors. 
Balanced accuracy is their mean. Valid rate is the fraction of schema-valid outputs returned within the per-sample runtime budget. 
LLM models are evaluated on the full benchmark, whereas LLM agents are evaluated on a balanced 500/500 subset.
}
\label{tab:main_results_all}
\end{table*}

Several patterns emerge from Table~\ref{tab:main_results_all}.
First, current systems struggle particularly to identify \texttt{not\_merge\_ready} snapshots.
Even with the input restricted to build-passing snapshots, the best Stage~3 LLM-model balanced accuracy is only \(36.0\%\), achieved by DeepSeek.
Qwen reaches \(24.1\%\), and Goedel and Kimina remain much lower. 
This indicates that \textsc{MathlibPR} probes a substantially harder capability than detecting syntax errors or compilation failure. 
The model must instead reason about the review norms by which Mathlib curates an integrated mathematical library.


Second, richer context yields only limited and uneven gains. 
Moving from Stage~1 to Stage~3 raises balanced accuracy for DeepSeek (\(18.1\% \to 36.0\%\)) and Qwen (\(15.8\% \to 24.1\%\)), 
but these gains are not matched by recognition of \texttt{not\_merge\_ready} snapshots. 
DeepSeek's Stage~3 NMR Recall is only \(2.3\%\), and Qwen's NMR Recall remains \(0.0\%\) across all three stages. 
Other systems do not even show monotone gains. 
Goedel's valid rate collapses from \(68.3\%\) at Stage~1 to \(13.5\%\) at Stage~3. 
Even when local code context is supplemented by diagnostics and PR intent, the reviewer-like judgment required by Mathlib is not captured well by current systems.


Third, access to repository state does not close the gap. 
LLM agents can inspect a read-only local checkout and use coding tools, but they still operate in the same overall difficulty regime. 
Codex achieves high MR Recall (\(92.0\%\) at Stage~3) and is operationally reliable (Stage~3 valid rate \(99.9\%\)), but its Stage~3 NMR Recall is only \(12.0\%\). 
Claude Code shows the opposite reliability profile, returning a valid output on only \(56.2\%\) of Stage~3 examples. 
In both cases, repository access does not help systems reliably identify \texttt{not\_merge\_ready} snapshots.

Behind these recall metrics, systems differ substantially in what they actually output. 
Table~\ref{tab:prediction_breakdown_all} reports the Stage~3 prediction distribution across all evaluated systems. 
We focus on Stage~3 because it provides the richest context and thus the clearest view of each system's residual behavior.
\begin{table*}[t]
\centering
\small
\setlength{\tabcolsep}{4.5pt}
\begin{tabular}{llcccc}
\toprule
Model & Eval.\ set & Pred.\ MR & Pred.\ NMR & Pred.\ uncertain & Invalid \\
\midrule
\multicolumn{6}{l}{\textit{LLM models, Stage~3}} \\
DeepSeek & Full & 68.9 & 2.1 & 21.9 & 7.1 \\
Qwen & Full & 48.3 & 0.0 & 48.7 & 3.0 \\
Goedel & Full & 1.5 & 0.2 & 11.8 & 86.5 \\
Kimina & Full & 0.0 & 11.3 & 0.0 & 88.7 \\
\midrule
  \multicolumn{6}{l}{\textit{LLM agents, Stage~3}} \\
  Codex & 500/500 & 87.5 & 8.3 & 4.1 & 0.1 \\
  Claude Code & 500/500 & 30.4 & 8.5 & 17.3 & 43.8 \\
\bottomrule
\end{tabular}
\caption{Stage~3 prediction breakdown for all evaluated systems. Values are percentages over the relevant evaluation set and may not sum to exactly 100 due to rounding.}
\label{tab:prediction_breakdown_all}
\end{table*}

The four states are distributed very unevenly across systems, often clustering on a single state. 
DeepSeek and Codex commit to \texttt{merge\_ready} on the large majority of examples while almost never committing to \texttt{not\_merge\_ready}. 
Qwen instead splits its outputs roughly evenly between \texttt{merge\_ready} and \texttt{uncertain}, again almost never committing to \texttt{not\_merge\_ready}. 
Even Codex, the most operationally 
reliable system, commits to \texttt{not\_merge\_ready} on only 
\(8.3\%\) of examples.

Goedel, Kimina, and Claude Code instead concentrate on invalid outputs. 
To understand these failures, we examined their Stage~3 invalid outputs. 
All of Goedel's invalid cases are prose-style analyses, and all of Kimina's are Lean code blocks. 
In those invalid cases, the models do not produce the required reviewer-JSON output at all.
Both Goedel and Kimina are trained for Lean theorem proving, and this pattern is consistent with a mismatch between their training objective and the structured review interface required by our benchmark.
Claude Code's invalid cases differ. 
Its finalized invalid outputs are mostly malformed near-complete JSON, while the raw logs also show a large volume of timeout-driven failures, with \(35.8\%\) of Stage~3 latest-attempt outputs exceeding the per-sample runtime budget. 
These failure modes therefore reflect interface or budget constraints rather than valid reviewer judgments. 
Representative invalid outputs are shown in Appendix~\ref{app:invalid-outputs}.

We also examine the effect of excluding never-merged PRs (i.e., PRs closed without being merged). Excluding these 799 negative snapshots changes AUROC and balanced accuracy by less than one percentage point across all systems and stages (Appendix~\ref{app:merged-only-results}).


\section{Analysis and Error Study}
\label{sec:analysis}

We now characterize when systems abstain via \texttt{uncertain}, and test whether they can recognize revision progress within the same PR.

\subsection{Abstention Patterns}
\label{sec:abstentions}

Abstention via \texttt{uncertain} is not a single uniform behavior. 
Table~\ref{tab:prediction_breakdown_all} shows that Stage~3 abstention rates range from \(0.0\%\) (Kimina) to \(48.7\%\) (Qwen), with nearly half of Qwen's outputs being abstentions. 
To understand what drives these abstentions, we report in Table~\ref{tab:uncertain_axis_categories} which rubric axes each system flags as concerning when it abstains (axes defined in 
Appendix~\ref{app:prompts-schema}).

\begin{table*}[t]
\centering
\small
\setlength{\tabcolsep}{4.5pt}
\begin{tabular}{lcccccccc}
\toprule
Model & Doc. & Naming & Local & File & Imports & Proof & API fit & Overlap \\
\midrule
\multicolumn{9}{l}{\textit{LLM models, Stage~3}} \\
DeepSeek & 75.8 & 3.6 & 1.1 & 0.5 & 1.4 & 28.3 & 2.4 & 2.6 \\
Qwen & 83.0 & 3.6 & 1.1 & 0.2 & 1.4 & 16.8 & 0.9 & 0.8 \\
Goedel & 69.1 & 17.1 & 23.1 & 5.1 & 17.7 & 50.2 & 22.7 & 17.1 \\
\midrule
\multicolumn{9}{l}{\textit{LLM agents, Stage~3}} \\
Codex & 75.6 & 58.5 & 34.1 & 0.0 & 31.7 & 58.5 & 0.0 & 29.3 \\
Claude Code & 86.1 & 43.4 & 46.2 & 5.8 & 8.7 & 28.9 & 24.3 & 5.8 \\
\bottomrule
\end{tabular}
\caption{Concern categories among valid \texttt{uncertain} outputs in Stage~3. Each cell reports the percentage of a system's valid \texttt{uncertain} outputs that flag the corresponding axis as \texttt{concern} or \texttt{blocker}. Categories are not mutually exclusive. Kimina is omitted because its Stage~3 run contains only one valid \texttt{uncertain} output.}
\label{tab:uncertain_axis_categories}
\end{table*}

The dominant axes also differ across systems, in ways that track their inspection capability. 
DeepSeek and Qwen abstain overwhelmingly on documentation (\(75.8\%\) and \(83.0\%\) of their abstentions, respectively), with proof readability a distant second. 
Both axes are visible from the diff alone, consistent with LLM models that judge from the prompt without further inspection. 
Goedel, pre-fine-tuned on Mathlib, also abstains heavily on documentation but additionally raises proof readability concerns on roughly half of its abstentions. 
This suggests it brings more internalized knowledge of Mathlib's proof style. 
The LLM agents behave differently. 
Codex and Claude Code spread their abstentions across more axes, including naming, local structure, imports, and overlap.
These axes benefit from inspecting the surrounding code rather than the diff in isolation. 
This contrast suggests that part of the abstention behavior reflects what each system can observe, not just what it judges to matter.

\subsection{Within-PR Comparison}
\label{sec:within_pr_pairs}

We next test whether systems can recognize revision progress inside the same PR. 
We construct pairs by taking an earlier build-passing snapshot and the final accepted snapshot from the same merged PR. 
For each pair, let \(s_{\mathrm{early}}\) and \(s_{\mathrm{final}}\) be the system's scalar merge-readiness scores for the two snapshots, as defined in the schema (Appendix~\ref{app:prompts-schema}). 
We score the pair as correct if \(s_{\mathrm{final}} > s_{\mathrm{early}}\), incorrect if \(s_{\mathrm{final}} < s_{\mathrm{early}}\), and half-correct if the two scores tie. 
Pairwise accuracy averages this 0/0.5/1 score over usable pairs, where a pair is usable if both snapshots received a schema-valid output. 
Pairwise accuracy thus measures whether the system ranks the accepted final snapshot above the earlier one, not whether it predicts the correct binary label. 
Table~\ref{tab:within_pr_pairs} reports this metric for LLM models. 
We restrict the analysis to LLM models because the LLM agents are evaluated on a balanced 500/500 subset, which yields fewer than 15 paired PRs per stage and is too small for stable pairwise estimates.



\begin{table*}[t]
\centering
\small
\setlength{\tabcolsep}{4.5pt}
\begin{tabular}{llrrrr}
\toprule
Model & Stage & Total pairs & Usable pairs & Pairwise Acc. & Mean $\Delta$ score \\
\midrule
\multicolumn{6}{l}{\textit{LLM models}} \\
DeepSeek & S1 & 3687 & 3563 & 38.0 & -0.091 \\
DeepSeek & S2 & 3687 & 3467 & 52.5 & 0.019 \\
DeepSeek & S3 & 3687 & 3101 & 51.1 & 0.006 \\
Qwen & S1 & 3687 & 3460 & 50.4 & 0.004 \\
Qwen & S2 & 3687 & 3442 & 51.3 & 0.009 \\
Qwen & S3 & 3687 & 3479 & 49.1 & -0.002 \\
Goedel & S1 & 3687 & 610 & 28.4 & -0.106 \\
Goedel & S2 & 3687 & 151 & 50.3 & 0.001 \\
Goedel & S3 & 3687 & 101 & 49.0 & -0.006 \\
Kimina & S1 & 3687 & 297 & 49.7 & -0.003 \\
Kimina & S2 & 3687 & 90 & 48.9 & -0.009 \\
Kimina & S3 & 3687 & 99 & 49.0 & 0.008 \\
\bottomrule
\end{tabular}
\caption{Within-PR pairwise analysis for LLM models on the full 
benchmark, restricted to merged PRs with both an earlier 
build-passing snapshot and the final accepted snapshot. Pairwise 
Acc.\ counts ties as half-correct. Mean $\Delta$ score is the 
average difference between the final and earlier snapshots' 
merge-readiness scores.}
\label{tab:within_pr_pairs}
\end{table*}

LLM models barely beat chance on this controlled comparison. 
DeepSeek improves from Stage~1 to Stage~2 but plateaus at Stage~3, remaining close to chance. 
Qwen stays at chance across all stages. 
Goedel and Kimina retain too few usable pairs for their pairwise numbers to be informative. 
This pairwise comparison strengthens the main result. 
Failing to distinguish \texttt{merge\_ready} from \texttt{not\_merge\_ready} across the full benchmark could in principle reflect only that build-passing PRs all look broadly similar to a given system. 
The within-PR comparison removes that confound, since both snapshots are from the same PR, and current models still cannot reliably tell which version was accepted.

To complement the preceding quantitative analyses, we conduct two manual audits of the underlying review evidence. We examine whether never-merged PRs have documented technical concerns visible in the evaluated snapshots, and whether correct \texttt{not\_merge\_ready} predictions identify concerns reflected in subsequent reviewer feedback. The results are reported in Appendix~\ref{app:manual-audits}.

\section{Discussion and Limitations}
\label{sec:discussion}

\paragraph{Limitations.}
The benchmark covers a selective slice of closed PRs. 
\texttt{not\_merge\_ready} snapshots require recoverable PR-side build-check evidence, while \texttt{merge\_ready} snapshots are recovered from accepted master-side commits. 
Combined with our temporal cutoff, this means that \textsc{MathlibPR} does not exhaustively represent all closed PR activity in Mathlib's history. 
The labels are derived from historical merge outcomes rather than independent assessments of code quality, and some never-merged PRs may have been closed for non-technical reasons such as timing, author abandonment, or reviewer availability.
A matched evaluation by Mathlib experts under the same evidence constraints remains future work and would help assess task ambiguity and the human–model gap.
Finally, LLM-agent evaluation is currently limited to a balanced 500/500 subset because of cost, which reduces direct comparability with full-benchmark model results and limits some downstream analyses such as the within-PR comparison in Section~\ref{sec:within_pr_pairs}.

\paragraph{Implications and future work.}
The goal of \textsc{MathlibPR} is not to replace Mathlib reviewers. Human review also supports community discussion, consensus-building, and contributor learning, which fully automated decisions could undermine. \textsc{MathlibPR} provides a benchmark for studying whether models can approximate part of the review signal that arises after code already passes formal checking, with predictions serving as advisory signals subject to human review rather than as a basis for automatic PR decisions.
We see several concrete directions for using and extending it. 
One is reviewer assistance and PR triage, where systems may help surface likely blockers or prioritize attention without making final decisions. 
Another is training reward models for merge-ready code. 
The within-PR pair structure provides a natural preference signal, since a capable system should prefer the final accepted snapshot over an earlier build-passing but not-yet-accepted snapshot from the same PR. 
The benchmark also admits a prospective extension over time. 
Because the current release is defined by a historical cutoff, later-closed PRs and currently open PRs can support forward-looking evaluations that reduce overfitting to absorbed repository history.

\section{Conclusion}
\label{sec:conclusion}

We introduced \textsc{MathlibPR}, the first benchmark for evaluating whether LLMs can predict Mathlib PR merge outcomes.
Built from real Mathlib4 PR histories, the benchmark turns historical maintainer decisions into a supervised signal that probes the gap between build-passing code and merge-ready code. 
We further proposed a three-stage evaluation protocol that progressively supplies review-relevant context, and used it to evaluate both LLM models and LLM agents. 
Across all evaluated systems, none reliably distinguishes \texttt{merge\_ready} from \texttt{not\_merge\_ready} snapshots, and neither richer context nor repository access closes this gap. 
As the LLM-assisted formal mathematics ecosystem moves from consuming Mathlib to contributing back to it, judging whether generated code is merge-ready becomes the next bottleneck. 
\textsc{MathlibPR} provides a step toward reviewer assistants and reward models that can help close this bottleneck.

\section*{Acknowledgments}
This work is supported in part by the US National Science Foundation under the awards III-2128019, SLES-2331904, and CAREER-2442098, the Commonwealth Cyber Initiative's Central Virginia Node under the award VV-1Q26-001, VV-1Q27-007, and VV-1Q27005, 4VA at UVA, a Cisco Faculty Research Award, and a Google Research Award.


{\small
\bibliography{bibliography}
}

\newpage
\appendix
\onecolumn
\section{Benchmark Access and Manifest}
\label{app:access-manifest}

\subsection{Access}
\label{app:access}

\textsc{MathlibPR} is hosted on Hugging Face at \url{https://huggingface.co/datasets/MathlibPR/MathlibPR}. 
The benchmark is derived from the public \texttt{leanprover-community/mathlib4} repository and the Mathlib contribution documentation, both distributed under Apache-2.0.
The derived benchmark and its evaluation harness are released under Apache-2.0 as well.
The release includes a Croissant metadata file with both core dataset metadata and Responsible AI fields, conforming to NeurIPS 2026 E\&D track requirements. 
Full field documentation, loading code, prompt materialization utilities, scoring scripts, and reconstruction scripts are maintained in the dataset card.

We additionally release 95\% confidence intervals for the main
evaluation metrics and paired stage comparisons, computed using
10,000 PR-level cluster bootstrap resamples to account for
dependencies among snapshots from the same PR.
These results, along with analyses by PR characteristics,
are available in our
\href{https://huggingface.co/datasets/MathlibPR/MathlibPR/tree/main/rebuttal_audit/statistics}
{released statistical analyses}.

\subsection{Manifest Schema Overview}
\label{app:manifest-schema}

Each \textsc{MathlibPR} example is represented by a compact source-manifest record that stores the snapshot's identity, provenance, and reconstruction metadata. 
Table~\ref{tab:manifest-fields} groups the manifest fields by role.
\begin{table}[h]
\centering
\small
\setlength{\tabcolsep}{5pt}
\begin{tabular}{llc}
\toprule
Field group & Example fields & Scope \\
\midrule
Identity & \texttt{sample\_id}, \texttt{pr\_number}, 
\texttt{target\_merged}, \texttt{snapshot\_role} & record \\
Snapshot reconstruction & \texttt{snapshot\_commit\_sha}, 
\texttt{diff\_base\_sha}, \texttt{diff\_source} & record \\
Selection provenance & \texttt{source\_commit\_seq}, 
\texttt{source\_commit\_sha}, \texttt{selection\_diff\_path} & record \\
Pairing metadata & \texttt{negative\_commit\_seq}, 
\texttt{final\_commit\_seq}, \texttt{last\_commit\_seq}, & record \\
& \texttt{final\_snapshot\_source}, 
\texttt{merge\_commit\_source} & \\
Diagnostics & \texttt{linter\_diagnostics}, 
\texttt{import\_diagnostics}, \texttt{location\_diagnostics}, & 
S2--3 \\
& \texttt{doc\_diagnostics}, \texttt{api\_diagnostics} & \\
PR intent & \texttt{pr\_title}, \texttt{pr\_description} & S3 \\
\bottomrule
\end{tabular}
\caption{Manifest field groups for a single \textsc{MathlibPR} 
record. The Scope column indicates whether a group is 
record-level metadata or stage-gated evidence shown to the system 
at the indicated stages. The dataset card lists the complete 
field set with per-field types and semantics.}
\label{tab:manifest-fields}
\end{table}

The identity fields uniquely identify the snapshot and carry its 
binary reference label \texttt{target\_merged} along with the 
snapshot type recorded in \texttt{snapshot\_role}, such as 
\texttt{first\_build\_success\_snapshot} or 
\texttt{final\_snapshot}. The snapshot reconstruction fields 
specify which repository state is evaluated and the base against 
which the cumulative diff is computed. The selection provenance 
fields record which PR-side event or commit produced the selected 
snapshot, supporting reproducibility audits. The pairing metadata 
fields link an early build-passing snapshot to the corresponding 
final accepted or final unmerged commit when applicable, and are 
used to assemble the within-PR pair manifest. The diagnostics 
fields hold the recovered or backfilled lint, import, location, 
documentation, and API check results that Stage~2 and Stage~3 
prompts incorporate as evidence. The PR intent fields hold the 
PR title and description, exposed only at Stage~3.

\subsection{Illustrative Record}
\label{app:manifest-example}

The following example shows a \texttt{not\_merge\_ready} record 
(\texttt{target\_merged = 0}) for the first build-passing 
snapshot of an unmerged PR. Some provenance, pairing, and 
diagnostics fields are omitted for brevity.

\begin{lstlisting}
{
  "sample_id": "prXXXXX_neg_first_build_success_snapshot_seqK_<sha12>",
  "pr_number": XXXXX,
  "target_merged": 0,
  "snapshot_role": "first_build_success_snapshot",
  "diff_source": "cumulative_pr_snapshot",
  "snapshot_commit_sha": "<snapshot_commit_sha>",
  "diff_base_sha": "<diff_base_sha>",
  "source_commit_seq": K,
  "selection_diff_path": "commits/XXXXX_K_<sha7>.diff",
  "final_snapshot_source": "last_pr_commit_unmerged",
  "pr_title": "<title or null>",
  "pr_description": "<description or null>"
}
\end{lstlisting}

For an accepted positive snapshot, \texttt{target\_merged} is 
\(1\), \texttt{snapshot\_role} is typically \texttt{final\_snapshot} 
or \texttt{single\_build\_success\_commit\_snapshot}, and 
\texttt{diff\_source} is \texttt{merged\_commit\_patch}. The 
release also includes a separate within-PR pair manifest with 
\texttt{pair\_id}, \texttt{earlier\_sample\_id}, and 
\texttt{final\_sample\_id} fields linking the earlier 
build-passing snapshot to the final accepted snapshot of the same 
PR.

\subsection{Agent Subset Sampling}
\label{app:subset-sampling}

We evaluate LLM agents on a balanced 500/500 subset rather 
than the full benchmark because the agent setting is 
substantially more expensive than prompt-only evaluation. The 
subset is sampled with a fixed random seed (\texttt{seed=42}) 
and contains 500 \texttt{merge\_ready} and 500 
\texttt{not\_merge\_ready} examples selected from the 
full benchmark.

We use separate released subsets for each stage, since the stage-specific benchmark inputs are materialized separately and the finalized agent-stage evaluations are tracked separately in the release. 
Stage~1 uses code-local context only. Stage~2 adds diagnostics. Stage~3 adds both diagnostics and PR title or description as stated intent. 
Subset memberships overlap substantially across stages but are not identical. 
The released benchmark therefore provides all six stage-by-agent subsets as separate dataset configs for direct reuse.

\section{Prompt Templates and Output Schema}
\label{app:prompts-schema}

\subsection{Prompt Design Overview}
\label{app:prompt-overview}

All stages share a common reviewer-oriented contract: given a build-passing Mathlib4 PR snapshot, the system must judge whether the snapshot is \texttt{merge\_ready} or \texttt{not\_merge\_ready}, or abstain via \texttt{uncertain}. 
The contract instructs the system to make a code-centric judgment, ignore CI status, authorship, and merge outcome, and refrain from external retrieval. 
Table~\ref{tab:staged-prompts} summarizes which evidence is available at each stage; the stage-specific materializations are given in the prompt templates below.
\begin{table}[h]
\centering
\begin{tabular}{lccccc}
\toprule
Stage & Diff & Changed files & Guidelines & Diagnostics & PR intent \\
\midrule
Stage~1 & \checkmark & \checkmark & \checkmark & -- & -- \\
Stage~2 & \checkmark & \checkmark & \checkmark & \checkmark & -- \\
Stage~3 & \checkmark & \checkmark & \checkmark & \checkmark & \checkmark \\
\bottomrule
\end{tabular}
\caption{
Staged prompting protocol for \textsc{MathlibPR}. All 
stages use the same reviewer contract and JSON output schema. 
Stage 1 provides code-local evidence and compact guideline digests. 
Stage 2 adds automated diagnostics. Stage 3 adds PR title and 
description as stated intent. The exact prompts are given below.
}
\label{tab:staged-prompts}
\end{table}

\subsection{Output Schema}
\label{app:output-schema}

All systems return a structured JSON review that is validated against a fixed schema. 
LLM models and LLM agents share the same core schema, with one additional required field (\texttt{repo\_checks\_used}) included for LLM agents to record the repository inspections used in reaching the judgment. 
We describe the top-level fields, the verdict semantics, the eight rubric axes, and the axis label values in turn.

\subsubsection{Top-Level Fields}
\label{app:top-level-fields}

Each output JSON object contains the following shared 
top-level fields:

\begin{itemize}
\item \texttt{verdict}: the system's overall judgment of the 
snapshot, taking values \texttt{merge\_ready}, 
\texttt{not\_merge\_ready}, or \texttt{uncertain}.
\item \texttt{p\_merge\_ready}: a scalar merge-readiness score, 
interpreted on \([0,1]\), that we use for ranking-based metrics 
such as AUROC and within-PR pairwise accuracy.
\item \texttt{overall\_confidence}: the system's stated 
confidence in its own verdict.
\item \texttt{axes}: an object containing per-axis assessments 
along the eight rubric axes defined in 
Appendix~\ref{app:axis-definitions}.
\item \texttt{top\_strengths}: a short list of strengths 
identified in the snapshot.
\item \texttt{top\_blockers}: a short list of issues that the 
system considers blockers to merge.
\item \texttt{minimal\_required\_changes}: a short list of 
changes the system would request before merge.
\item \texttt{other\_concerns}: additional concerns that do 
not rise to the level of top blockers.
\end{itemize}

The complete core schema used for LLM model runs is given below.
  \begin{lstlisting}
  {
    "type": "object",
    "additionalProperties": false,
    "properties": {
      "verdict": {
        "type": "string",
        "enum": ["merge_ready", "not_merge_ready", "uncertain"]
      },
      "p_merge_ready": {"type": "number"},
      "overall_confidence": {"type": "number"},
      "axes": {
        "type": "object",
        "additionalProperties": false,
        "properties": {
          "naming_style": {"$ref": "#/$defs/axis"},
          "documentation": {"$ref": "#/$defs/axis"},
          "local_structure": {"$ref": "#/$defs/axis"},
          "file_placement": {"$ref": "#/$defs/axis"},
          "imports_dependencies": {"$ref": "#/$defs/axis"},
          "proof_readability": {"$ref": "#/$defs/axis"},
          "api_library_fit": {"$ref": "#/$defs/axis"},
          "repository_overlap_generality": {"$ref": "#/$defs/axis"}
        },
        "required": [
          "naming_style",
          "documentation",
          "local_structure",
          "file_placement",
          "imports_dependencies",
          "proof_readability",
          "api_library_fit",
          "repository_overlap_generality"
        ]
      },
      "top_strengths": {
        "type": "array",
        "items": {"type": "string"}
      },
      "top_blockers": {
        "type": "array",
        "items": {"type": "string"}
      },
      "minimal_required_changes": {
        "type": "array",
        "items": {"type": "string"}
      },
      "other_concerns": {
        "type": "array",
        "items": {"type": "string"}
      }
    },
    "required": [
      "verdict",
      "p_merge_ready",
      "overall_confidence",
      "axes",
      "top_strengths",
      "top_blockers",
      "minimal_required_changes",
      "other_concerns"
    ],
    "$defs": {
      "axis": {
        "type": "object",
        "additionalProperties": false,
        "properties": {
          "label": {
            "type": "string",
            "enum": ["good", "concern", "blocker", "unknown"]
          },
          "confidence": {"type": "number"},
          "evidence": {
            "type": "array",
            "items": {"type": "string"},
            "maxItems": 3
          }
        },
        "required": ["label", "confidence", "evidence"]
      }
    }
  }
  \end{lstlisting}
For LLM agents, the schema additionally requires \texttt{repo\_checks\_used}, a list of inspected files, paths, or search patterns used during read-only repository 
inspection:

\begin{lstlisting}
"repo_checks_used": {
  "type": "array",
  "items": {"type": "string"}
}
\end{lstlisting}

with \texttt{"repo\_checks\_used"} appended to the top-level 
\texttt{required} list.

\subsubsection{Verdict Semantics}
\label{app:verdict-semantics}

The \texttt{verdict} field has three possible values:

\begin{itemize}
\item \texttt{merge\_ready}: the system judges that the 
snapshot is ready to be accepted into Mathlib4.
\item \texttt{not\_merge\_ready}: the system judges that the 
snapshot should not be accepted in its current form.
\item \texttt{uncertain}: the system declines to issue either 
judgment. We treat \texttt{uncertain} as a substantive 
abstention rather than as an invalid output.
\end{itemize}

The reference label is binary, 
so \texttt{uncertain} predictions are counted as errors for 
class-recall and balanced-accuracy calculations.

\subsubsection{Axis Definitions}
\label{app:axis-definitions}

The \texttt{axes} object contains one assessment for each of 
the following eight rubric axes:

\begin{itemize}
\item \texttt{naming\_style}: whether declaration names, local 
names, notation, and naming patterns follow Mathlib naming and 
style conventions.
\item \texttt{documentation}: whether the snapshot provides 
adequate docstrings, module-level documentation, comments, and 
other explanatory text expected for public-facing Mathlib 
contributions.
\item \texttt{local\_structure}: whether the code is well 
organized within the touched files, including declaration 
ordering, namespace and section structure, and local 
maintainability.
\item \texttt{file\_placement}: whether the changed declarations 
are located in the appropriate file, module, or part of the 
repository rather than belonging elsewhere.
\item \texttt{imports\_dependencies}: whether imports and 
dependencies are appropriate and minimal, avoiding unnecessary 
coupling or misplaced dependency choices.
\item \texttt{proof\_readability}: whether proofs are clear, 
idiomatic, and maintainable, rather than opaque, brittle, or 
unnecessarily difficult to review.
\item \texttt{api\_library\_fit}: whether the contribution 
integrates well with the surrounding Mathlib API, abstraction 
boundaries, and existing library design patterns.
\item \texttt{repository\_overlap\_generality}: whether the 
proposed declarations are sufficiently general and not 
redundant with existing or more general results in the 
repository.
\end{itemize}

Each axis assessment consists of three fields: \texttt{label}, 
\texttt{confidence}, and \texttt{evidence}. The 
\texttt{evidence} field is a list of up to three short strings, 
each citing a specific concern or strength observed in the 
snapshot.

\subsubsection{Axis Label Semantics}
\label{app:axis-label-semantics}

Each axis label takes one of four values:

\begin{itemize}
\item \texttt{good}: the system finds no material concern on 
this axis.
\item \texttt{concern}: the system identifies an issue on this 
axis that may merit reviewer attention but is not by itself a 
blocker.
\item \texttt{blocker}: the system identifies an issue on this 
axis that it considers a blocker to merge.
\item \texttt{unknown}: the available evidence is insufficient 
for the system to make a confident axis-level judgment. We 
reserve \texttt{unknown} for axis-level evidence insufficiency, 
distinct from the verdict-level abstention \texttt{uncertain}.
\end{itemize}

\subsection{Prompt Templates for LLM Models}
\label{app:llm-model-prompts}

The LLM model setting follows the standard chat-completion message format with two messages per evaluation. 
The system message fixes the reviewer contract and is identical across all stages and samples. 
The user message is stage-specific and instantiated per sample by filling in the diff, changed files, diagnostics, and PR intent placeholders. 
We give the system message once below, then show the Stage~1 user message in full, followed by the incremental additions for Stage~2 and Stage~3. 

\paragraph{System prompt.}
\noindent
\begin{lstlisting}
You are reviewing a Mathlib4 PR snapshot for code-centric merge-readiness.

Use only the supplied context. Ignore CI/build status, workflow or bors state,
commit order, timeline, author/reviewer identity, and eventual merge outcome.
Do not use repository access, web search, or external tools.

Rubric axes:
- naming_style
- documentation
- local_structure
- file_placement
- imports_dependencies
- proof_readability
- api_library_fit
- repository_overlap_generality

Axis labels: good, concern, blocker, unknown.
Use unknown only for axes with insufficient evidence.
Use uncertain only for the final verdict.

Return only the required JSON object. Keep evidence short and concrete.
\end{lstlisting}

\paragraph{Stage 1 user prompt.}
\noindent
\begin{lstlisting}
<task>
Estimate whether this snapshot is merge-ready for Mathlib4 using only the
supplied evidence.
</task>

<context>
<diff_chunks>
{diff_chunks}
</diff_chunks>

<naming_guidelines_markdown>
{naming_md}
</naming_guidelines_markdown>

<style_guidelines_markdown>
{style_md}
</style_guidelines_markdown>

<documentation_guidelines_markdown>
{doc_md}
</documentation_guidelines_markdown>

<changed_files_complete>
{changed_files_complete}
</changed_files_complete>

<first_order_import_files>
{first_order_import_files}
</first_order_import_files>
</context>

<instructions>
1. Use the rubric axes as the checklist.
2. Use the naming/style/doc digests as authoritative written guidance.
3. Use changed files and first-order imports to judge local conventions and
   code quality.
4. Use unknown instead of guessing when evidence is insufficient.
5. Output only the required JSON object.
</instructions>
\end{lstlisting}

\paragraph{Stage 2 user prompt.}

The Stage~2 user prompt extends the Stage~1 prompt by inserting 
a \texttt{<diagnostics>} block inside \texttt{<context>}, and 
replacing the third item in \texttt{<instructions>} with one 
that mentions diagnostics. The new block is:

\begin{lstlisting}
<diagnostics>
{linter_diagnostics}
{import_diagnostics}
{location_diagnostics}
{doc_diagnostics}
{api_diagnostics}
</diagnostics>
\end{lstlisting}

The updated instruction reads:

\begin{lstlisting}
3. Treat diagnostics as evidence, not as automatically decisive.
\end{lstlisting}

All other content is identical to Stage~1.

\paragraph{Stage 3 user prompt.}

The Stage~3 user prompt extends the Stage~2 prompt by adding a 
\texttt{<pr\_intent>} block at the end of \texttt{<context>}, and 
replacing the third instruction with one that mentions PR 
intent. The new block is:

\begin{lstlisting}
<pr_intent>
title: {pr_title}
description:
{pr_description}
</pr_intent>
\end{lstlisting}

The updated instruction reads:

\begin{lstlisting}
3. Treat the PR title/description only as stated intent; code and diagnostics control the judgment.
\end{lstlisting}

All other content is identical to Stage~2.

\subsection{Prompt Templates for LLM Agents}
\label{app:llm-agent-prompts}

The stage-specific user-message templates are shared between 
Codex and Claude Code. 
The system messages share a common reviewer contract; Claude Code additionally inlines backend-specific tool and output-format rules and the agent JSON schema. 
The Codex backend receives the schema out-of-band via the structured-output API rather than inline in the system message.

\paragraph{Codex system message.}
\noindent
\begin{lstlisting}
You are reviewing a mathlib4 PR snapshot for code-centric merge-readiness.

Use the supplied context and read-only repository inspection only. 
Ignore CI/build status, workflow or bors state, commit order, timeline, author/reviewer identity, and eventual merge outcome. 
Do not use web search, GitHub or mathlib4 PR pages, git history, remote APIs, or external tools beyond the allowed read-only repository checks. 
Do not fetch network content with a browser, curl, wget, Python requests, or similar tools.

Rubric axes:
- naming_style
- documentation
- local_structure
- file_placement
- imports_dependencies
- proof_readability
- api_library_fit
- repository_overlap_generality

Axis labels: good, concern, blocker, unknown.
Use unknown only for axes with insufficient evidence.
Use uncertain only for the final verdict.

Return only the required JSON object. Keep evidence short and concrete.
\end{lstlisting}

\paragraph{Claude Code system message.}
The Claude Code system message begins with the same body as the 
Codex system message above and appends the following 
backend-specific section:

\begin{lstlisting}
Additional rules for this Claude Code run:
- Inspect the repository snapshot using the provided read-only tools (Read, Glob, and restricted Bash).
- Do not edit files, create files, install dependencies, or access any network or web resources.
- Do not open GitHub, mathlib4 PR pages, issue pages, or any remote webpage or API.
- Do not use git history commands (git log, git blame, git show, etc.).
- Return ONLY the final JSON object matching the schema below. No markdown, no explanation, no code fences.

Required output JSON schema:
<!-- the inlined schema is identical to the agent schema given in Appendix B.2 -->
\end{lstlisting}

\paragraph{Stage 1 user message.}
The Stage~1 user message is shared between Codex and Claude Code. 
Two elements are agent-specific: the \texttt{<task>} block frames the read-only repository as additional evidence beyond the supplied local context, and the \texttt{<repo\_workflow>} block (paste verbatim in Appendix~\ref{app:repo-workflow}) prescribes how the system should use this access. 
We replace the inlined \texttt{<repo\_workflow>} body with a placeholder here to avoid duplication.

\begin{lstlisting}
<task>
Estimate whether this snapshot is merge-ready for mathlib4.
You are inside a read-only repository snapshot corresponding to this PR snapshot.
Use the repository only to gather code-centric evidence that is not visible from the
provided local context.
</task>

<context>
<diff_chunks>
{diff_chunks}
</diff_chunks>

<naming_guidelines_markdown>
{naming_md}
</naming_guidelines_markdown>

<style_guidelines_markdown>
{style_md}
</style_guidelines_markdown>

<documentation_guidelines_markdown>
{doc_md}
</documentation_guidelines_markdown>

<changed_files_complete>
{changed_files_complete}
</changed_files_complete>

<first_order_import_files>
{first_order_import_files}
</first_order_import_files>
</context>

<repo_workflow>
<!-- see Appendix B.5 -->
</repo_workflow>

<instructions>
Return only the required JSON object.
If desired, populate repo_checks_used with inspected paths or search patterns.
</instructions>
\end{lstlisting}

\paragraph{Stage 2 user message.}
Stage~2 reuses the Stage~1 user message, inserts the following 
\texttt{<diagnostics>} block inside \texttt{<context>}, and 
replaces the Stage~1 \texttt{<repo\_workflow>} block with the 
Stage~2 variant in Appendix~\ref{app:repo-workflow}. The 
\texttt{<instructions>} block is unchanged.

\begin{lstlisting}
<diagnostics>
{linter_diagnostics}
{import_diagnostics}
{location_diagnostics}
{doc_diagnostics}
{api_diagnostics}
</diagnostics>
\end{lstlisting}

\paragraph{Stage 3 user message.}
Stage~3 reuses the Stage~2 user message, inserts the following 
\texttt{<pr\_intent>} block inside \texttt{<context>}, and 
replaces the Stage~2 \texttt{<repo\_workflow>} block with the 
Stage~3 variant in Appendix~\ref{app:repo-workflow}. The 
\texttt{<instructions>} block is unchanged.

\begin{lstlisting}
<pr_intent>
title: {pr_title}
description:
{pr_description}
</pr_intent>
\end{lstlisting}

\subsection{Repository Workflow Block}
\label{app:repo-workflow}

Codex and Claude Code share the same \texttt{<repo\_workflow>} 
templates. The block varies by stage because Stage~2 adds 
diagnostics as evidence and Stage~3 additionally introduces PR 
intent as stated scope.
\noindent
\paragraph{Stage 1 \texttt{<repo\_workflow>} block.}
\noindent
\begin{lstlisting}
<repo_workflow>
1. Start from the provided diff and changed files.
2. Inspect touched declarations and nearby declarations in the same files.
3. Search the repository snapshot for:
   - possible duplicate or more general existing results
   - naming analogues in the same namespace or nearby file families
   - plausible file/module homes for the changed declarations
4. Use read-only repository tools only.
5. Do not edit files or run build, test, lint, or CI-like commands.
6. Do not use git history, workflow metadata, internet access, GitHub/mathlib4 PR pages, or any remote API.
7. Use unknown instead of guessing when repo-wide evidence remains insufficient.
</repo_workflow>
\end{lstlisting}

\paragraph{Stage 2 \texttt{<repo\_workflow>} block.}
\noindent
\begin{lstlisting}
<repo_workflow>
1. Start from the provided diff, changed files, and diagnostics.
2. Inspect touched declarations and nearby declarations in the same files.
3. Search the repository snapshot for:
   - possible duplicate or more general existing results
   - naming analogues in the same namespace or nearby file families
   - plausible file/module homes for the changed declarations
4. Treat diagnostics as evidence, but verify them against the code and repository context.
5. Use read-only repository tools only.
6. Do not edit files or run build, test, lint, or CI-like commands beyond the supplied diagnostics.
7. Do not use git history, workflow metadata, internet access, GitHub/mathlib4 PR pages, or any remote API.
8. Use unknown instead of guessing when repo-wide evidence remains insufficient.
</repo_workflow>
\end{lstlisting}

\paragraph{Stage 3 \texttt{<repo\_workflow>} block.}
\noindent
\begin{lstlisting}
<repo_workflow>
1. Start from the provided diff, changed files, diagnostics, and PR intent.
2. Use PR title/description only to interpret intended scope and expected API/documentation surface.
3. Inspect touched declarations and nearby declarations in the same files.
4. Search the repository snapshot for:
   - possible duplicate or more general existing results
   - naming analogues in the same namespace or nearby file families
   - plausible file/module homes for the changed declarations
5. Treat diagnostics as evidence, but verify them against the code and repository context.
6. Use read-only repository tools only.
7. Do not edit files or run build, test, lint, or CI-like commands beyond the supplied diagnostics.
8. Do not use git history, workflow metadata, internet access, GitHub/mathlib4 PR pages, or any remote API.
9. Use unknown instead of guessing when repo-wide evidence remains insufficient.
</repo_workflow>
\end{lstlisting}

\section{Runtime, Tool Restrictions, and Validity Handling}
\label{app:runtime-validity}

\subsection{Fresh-Session Protocol}
\label{app:fresh-session}

Each evaluation example is processed in a fresh per-sample 
session. No state, history, or memory is carried across 
examples. This ensures that each judgment is grounded only in 
the supplied prompt, the read-only repository checkout, and 
the per-sample diagnostics or PR intent, with no leakage from 
prior samples in the same run.

\paragraph{Codex.}
Each sample is evaluated by a separate \texttt{codex exec} 
subprocess launched by the Python harness. The subprocess is 
invoked with \texttt{--ephemeral}, so no prior conversation 
thread is resumed and no model-side session state persists 
across samples. The stage-specific prompt is passed on 
standard input, and the working directory is a freshly 
materialized detached checkout for that sample. Harness-level 
resume affects only which sample IDs are skipped and does not 
reuse any model-side history.

\paragraph{Claude Code.}
Each sample is evaluated by a separate \texttt{claude -p} 
subprocess launched by the Python harness. The subprocess is 
invoked with \texttt{--no-session-persistence}, so no prior 
Claude session is resumed and no conversation history carries 
across samples. The stage-specific user prompt is passed on 
standard input, and the sample checkout is exposed via 
\texttt{--add-dir} and used as the subprocess working 
directory. As with Codex, harness-level resume affects only 
which examples are rerun and does not reuse model-side state.

\subsection{Allowed Inspection Tools}
\label{app:allowed-tools}

For reproducibility, we record the read-only inspection tools 
permitted in each agent setting. All other tools, including 
file editing, network access, and remote API calls, are 
disabled.

\paragraph{Codex.}
Codex runs combine the \texttt{read-only} sandbox with a 
locked shell profile that sets 
\texttt{shell\_environment\_policy.inherit=none} and replaces 
\texttt{PATH} with a temporary directory containing the 
following allowlisted binaries: \texttt{awk}, \texttt{bash}, 
\texttt{cat}, \texttt{cut}, \texttt{dirname}, \texttt{env}, 
\texttt{find}, \texttt{grep}, \texttt{head}, \texttt{ls}, 
\texttt{pwd}, \texttt{realpath}, \texttt{rg}, \texttt{sed}, 
\texttt{sh}, \texttt{sort}, \texttt{stat}, \texttt{tail}, 
\texttt{tr}, \texttt{uname}, \texttt{wc}, and \texttt{xargs}. 
Write tools, \texttt{git}, \texttt{curl}, \texttt{wget}, and 
\texttt{python} are excluded from this set, and any remaining 
write attempt is additionally blocked by the sandbox.

\paragraph{Claude Code.}
Claude Code is restricted by an explicit CLI allowlist. The 
allowed tools are \texttt{Read}, \texttt{Glob}, and a fixed 
set of read-only Bash command patterns covering 
\texttt{awk}, \texttt{cat}, \texttt{cut}, \texttt{dirname}, 
\texttt{find}, \texttt{grep}, \texttt{head}, \texttt{ls}, 
\texttt{pwd}, \texttt{realpath}, \texttt{rg}, \texttt{sed}, 
\texttt{sort}, \texttt{stat}, \texttt{tail}, \texttt{tr}, 
\texttt{wc}, and \texttt{xargs}. Edit/write tools, git-history 
commands, web or network access, and MCP tools are not 
permitted.

\subsection{Codex Runtime Configuration}
\label{app:codex-runtime}

We evaluate Codex via the Codex CLI 
(\texttt{codex exec}) backed by GPT-5.4. The runtime 
configuration is shared across stages, with only the dataset 
path, stage identifier, and selected sample set varying by 
run.

\paragraph{Configuration.}
\begin{itemize}
\item Model name: \texttt{gpt-5.4}
\item Session isolation: fresh per-sample subprocess
\item Sandbox mode: \texttt{read-only}
\item Structured output mode: JSON output constrained by an 
explicit schema
\item Reasoning configuration: \texttt{model\_reasoning\_effort="none"}
\end{itemize}

\paragraph{Invocation command.}
Each sample is invoked through a Python harness that wraps 
\texttt{codex exec} with a locked shell profile 
(\texttt{codex\_exec\_locked.sh}) constraining \texttt{PATH} 
to an allowlisted read-only tool directory in Appendix~\ref{app:allowed-tools}. The effective per-sample 
invocation is shown below.

\begin{lstlisting}[language=bash]
codex exec \
  -c `shell_environment_policy.inherit=none' \
  -c `shell_environment_policy.set={PATH="<tmp_locked_tools_dir>"}' \
  -c `model_reasoning_effort="none"' \
  --model gpt-5.4 \
  --sandbox read-only \
  --output-schema <schema_path> \
  --output-last-message <last_message_path> \
  --json \
  --color never \
  --ephemeral \
  -C <repo_root> \
  -
\end{lstlisting}

Here \texttt{<schema\_path>} and 
\texttt{<last\_message\_path>} are per-sample artifact paths, 
\texttt{<repo\_root>} is the detached checkout for the sampled 
snapshot, and the final \texttt{-} indicates that the prompt 
is supplied on standard input. The per-sample timeout cap was 
\texttt{900} seconds. In the final runs no completed Codex 
sample exceeded \texttt{300} seconds, so this cap was 
non-binding in practice.

\subsection{Claude Code Runtime Configuration}
\label{app:claude-runtime}

We evaluate Claude Code via the Claude Code CLI backed by 
\texttt{claude-sonnet-4-6}. The runtime configuration is 
shared across stages, with only the dataset path, stage 
identifier, and selected sample set varying by run.

\paragraph{Configuration.}
\begin{itemize}
\item Model name: \texttt{claude-sonnet-4-6}
\item Session isolation: fresh per-sample subprocess
\item Output format: \texttt{json}
\item Allowed tools: \texttt{Read}, \texttt{Glob}, and a 
restricted Bash subset (Appendix~\ref{app:allowed-tools})
\item Timeout: \texttt{300} seconds per sample
\end{itemize}

\paragraph{Invocation command.}
Each sample launches the following \texttt{claude -p} 
subprocess.

\begin{lstlisting}[language=bash]
claude -p \
  --model claude-sonnet-4-6 \
  --output-format json \
  --system-prompt "<Appendix B.4 Claude system message>" \
  --allowed-tools \
    "Read,Glob,\
Bash(cat:*),Bash(find:*),Bash(grep:*),Bash(head:*),Bash(ls:*),\
Bash(pwd),Bash(realpath:*),Bash(sed:*),Bash(sort:*),Bash(stat:*),\
Bash(tail:*),Bash(wc:*),Bash(xargs:*),Bash(awk:*),Bash(rg:*),\
Bash(tr:*),Bash(cut:*),Bash(dirname:*)" \
  --dangerously-skip-permissions \
  --no-session-persistence \
  --add-dir <repo_root>
\end{lstlisting}

Here \texttt{<repo\_root>} is the detached checkout for the 
sampled snapshot, and the stage-specific user prompt is 
supplied on standard input. The \texttt{300}-second cap 
reflects a compute budget rather than a benchmark requirement.

\subsection{Schema Validation and Invalid Outputs}
\label{app:invalid-outputs}

System outputs are parsed and validated against the JSON 
schema in Appendix~\ref{app:output-schema}. 
An output is  treated as invalid if it cannot be parsed as JSON, fails  schema validation, omits required fields, or exceeds the per-sample runtime budget. 
Invalid outputs are counted as errors for class-recall and balanced-accuracy calculations.

\paragraph{Parsing.}
The LLM model harness uses a permissive extraction pass before validation, since some providers emit fenced JSON, \texttt{<think>} traces, or near-schema variants such as rubric axes emitted at the top level rather than under \texttt{axes}. 
The harness extracts a candidate JSON payload, applies a small normalization step for these recurrent near-schema forms, and then validates against the schema. 
Codex and Claude Code use a stricter validator that operates on the agent's final returned message and does not strip markdown fences, remove \texttt{<think>} content, or issue a repair prompt after validation failure. 
Beyond plain JSON syntax errors, outputs are also marked invalid when they violate the schema semantically, including missing required fields, invalid enum values, wrong field types, malformed axis objects, or, for agent runs, omission of the required \texttt{repo\_checks\_used} field.

\paragraph{Representative invalid outputs.}
We provide one representative example per dominant failure 
mode observed in Section~\ref{sec:main}.

\textit{Goedel: prose-style analysis.}
This output is invalid because it never produces a 
schema-valid JSON review and instead returns free-form 
analysis.

\begin{lstlisting}
### Step 1: Understand the Problem

The problem involves evaluating whether a given Lean 4 code snippet
is merge-ready for the mathlib4 repository. The evaluation is based
on several axes, including naming style, documentation, local
structure, file placement, imports and dependencies, proof
readability, API and library fit, and repository overlap and
generality.

### Step 2: Analyze the Code
[...]
\end{lstlisting}

\textit{Kimina: Lean code block.}
This output is invalid because it returns a fenced Lean block 
rather than a review JSON object.

\begin{lstlisting}
```lean4
import Mathlib
import Mathlib.Algebra
import Mathlib.Data
import Mathlib.Tactic

/-!
# Mathlib4 merge-readiness analysis
[...]
-/
```
\end{lstlisting}

\textit{Claude Code: malformed near-complete JSON.}
This output is invalid because it is almost a complete review 
JSON but contains a later syntax error 
(\texttt{Expecting `,' delimiter}) and cannot be parsed. The 
truncation below shows only the well-formed prefix.

\begin{lstlisting}
{
"verdict": "not_merge_ready",
"p_merge_ready": 0.3,
"overall_confidence": 0.75,
"axes": {
"naming_style": {"label": "concern", "confidence": 0.85, ...},
"documentation": {"label": "concern", "confidence": 0.8, ...},
"local_structure": {"label": "concern", "confidence": 0.9, ...},
[...]
},
"repo_checks_used": ["Mathlib/NumberTheory/LucasLehmer.lean ...", ...]
}
[...]
\end{lstlisting}

\subsection{Timeout Policy}
\label{app:timeout}

Each sample is given a fixed runtime budget. Outputs 
exceeding this budget are terminated and counted as invalid 
for metric purposes.

\begin{itemize}
\item Codex per-sample timeout: \texttt{900}\,s
\item Claude Code per-sample timeout: \texttt{300}\,s
\item LLM model per-sample timeout: \texttt{180}\,s
\end{itemize}

When a Codex or Claude Code subprocess times out, the 
harness records the sample as invalid immediately. Partial 
output is not recovered and re-validated after the timeout 
fires. For LLM models, the timeout applies at the 
provider-request level. If the request times out, the 
harness receives no parseable final output and records the 
sample as failed unless a later request retry succeeds.

The final runs used different enforced timeout caps across 
backends. Codex was configured at 900\,s, but this cap was 
non-binding in practice. No completed Codex sample exceeded 
300\,s. Claude Code used an enforced 300\,s cap for compute 
budget reasons rather than as a benchmark-specific 
restriction.

\subsection{Retry and Repair Policy}
\label{app:retry-repair}

The LLM model harness retries provider requests up to three times for transient failures. 
If a response is returned but fails JSON parsing or schema validation, the harness may additionally issue up to 2 repair retries that re-prompt the model with the validation error and the previous invalid response, instructing it to return only the required JSON object. 
The agent harnesses do not implement an analogous repair pass. 
For Codex and Claude Code, any self-correction must occur within the agent's own single run, and a timeout or invalid final output is recorded immediately.

\subsection{Compute Resources}
\label{app:compute-resources}

Dataset construction, prompt materialization, schema validation, and metric computation run on CPU workers and require no local accelerator. 
LLM inference for the open-weight models was served locally with vLLM, and the LLM agents (Codex and Claude Code) were served by their respective backends, so backend-side accelerator resources for the agents are external to this paper. 
Each sample is processed in a fresh subprocess with the per-sample timeout caps reported in Appendix~\ref{app:timeout}.

\paragraph{LLM models.}
The 8B-scale models, Qwen3-8B and Kimina-Prover-Distill-8B, 
were served on a single H100. Each stage took approximately 
15 hours for Qwen and 24 hours for Kimina over the full 
benchmark. The 32B-scale models, DeepSeek-R1-Distill-Qwen 
and Goedel-Prover-V2-32B, were served on two A100 (80\,GB) 
GPUs each. Each stage took approximately 48 to 72 hours per 
model over the full benchmark.

\paragraph{LLM agents.}
The Codex and Claude Code agent runs were orchestrated from a 
CPU host, with model inference served remotely through the 
respective CLI backends. For Codex, each stage took 
approximately 30 hours over the balanced 500/500 subset. For 
Claude Code, each stage took approximately 80 hours over the 
balanced 500/500 subset.

\section{Input Packaging and Truncation}
\label{app:input-packaging}

This appendix records how staged prompt inputs are materialized 
from the manifest record and the repository checkout. The 
materialization is deterministic. Given a fixed manifest record 
and a fixed Mathlib4 checkout at the target snapshot, the 
resulting Stage~1, Stage~2, and Stage~3 prompts are 
reproducible.

\subsection{Guideline Digests}
\label{app:guideline-digests}

The final runs use three compact static guideline digests, 
one each for naming, style, and documentation, rather than the 
full contribution pages. These digests are shared across all 
samples and do not vary by sample. They are reduced markdown 
reference files derived from the corresponding Mathlib 
contribution guides\footnote{See 
\url{https://leanprover-community.github.io/contribute/naming.html}, 
\url{https://leanprover-community.github.io/contribute/style.html}, and 
\url{https://leanprover-community.github.io/contribute/doc.html}.}, with the reduction performed offline 
before evaluation rather than dynamically during prompt 
construction.

The final digest sizes are modest. The naming digest is 116 
lines (4{,}285 characters), the style digest is 104 lines 
(4{,}334 characters), and the documentation digest is 68 
lines (2{,}414 characters). In the reported runs, these digest 
blocks are not further truncated.

\subsection{Changed-File and Import Context}
\label{app:file-import-context}

The \texttt{changed\_files\_complete} block contains the 
snapshot versions of the Lean files changed between 
\texttt{diff\_base\_sha} and the target snapshot. Small files 
are included in full. Large files are reduced to the file 
header, touched declarations, and local windows around each 
changed hunk, with truncation surfaced by a 
\texttt{[TRUNCATED ...]} marker.

The \texttt{first\_order\_import\_files} block contains the 
modules named in the top-level \texttt{import} lines of the 
changed files. It is therefore a one-hop import context 
rather than a declaration-level dependency closure. Imported 
files are also truncated when necessary.

\subsection{Diagnostics Packaging}
\label{app:diagnostics-packaging}

Stage~2 and Stage~3 supply five diagnostics fields 
(\texttt{linter\_diagnostics}, \texttt{import\_diagnostics}, 
\texttt{location\_diagnostics}, \texttt{doc\_diagnostics}, 
\texttt{api\_diagnostics}), inserted in a fixed order inside 
a single \texttt{<diagnostics>} block. Recovered check-run 
records are rendered as short textual entries; build- and 
workflow-oriented checks are excluded. For accepted snapshots 
lacking recovered historical diagnostics, the release 
provides a locally backfilled lint summary, marked by a 
header such as \texttt{source=local\_lint\_style} so it is 
distinguishable from a recovered record.

\subsection{Length Limits}
\label{app:length-limits}

Input packaging uses fixed character caps rather than a 
sample-level token budget: 24{,}000 characters for 
\texttt{diff\_chunks}, 12{,}000 for 
\texttt{changed\_files\_complete}, 3{,}000 for 
\texttt{first\_order\_import\_files}, and 8{,}000 for each 
diagnostics field. When a block exceeds its cap, truncation 
preserves the beginning and appends a 
\texttt{[TRUNCATED ...]} marker.

\section{Evaluation Excluding Never-Merged PRs}
\label{app:merged-only-results}

We additionally examine the effect of excluding never-merged
PRs.
This removes 799 negative snapshots, leaving 11,409
positive and 3,687 negative snapshots from eventually
merged PRs. For LLM agents, we apply the same filtering
to their original evaluation subsets.

Table~\ref{tab:merged-only-results} compares performance
before and after filtering. AUROC and balanced accuracy
change by less than one percentage point across all systems
and stages, suggesting that our main findings are not
driven by never-merged PRs.

\begin{table}[t]
\centering
\small
\setlength{\tabcolsep}{7pt}
\begin{tabular}{llcc}
\toprule
Model & Stage & AUROC & Bal. Acc. \\
\midrule
\multicolumn{4}{l}{\emph{LLM models}} \\
DeepSeek & S1 & 40.5 $\to$ 40.8 & 18.1 $\to$ 18.1 \\
         & S2 & 51.4 $\to$ 52.2 & 32.4 $\to$ 32.4 \\
         & S3 & 52.4 $\to$ 53.0 & 36.0 $\to$ 36.1 \\
Qwen     & S1 & 49.9 $\to$ 49.8 & 15.8 $\to$ 15.8 \\
         & S2 & 49.7 $\to$ 49.2 & 19.3 $\to$ 19.3 \\
         & S3 & 48.4 $\to$ 47.9 & 24.1 $\to$ 24.1 \\
Goedel   & S1 & 30.0 $\to$ 29.2 & 0.2 $\to$ 0.2 \\
         & S2 & 51.3 $\to$ 52.2 & 1.4 $\to$ 1.4 \\
         & S3 & 53.1 $\to$ 52.8 & 1.0 $\to$ 0.9 \\
Kimina   & S1 & 49.5 $\to$ 49.5 & 9.3 $\to$ 9.6 \\
         & S2 & 51.9 $\to$ 51.7 & 4.1 $\to$ 4.4 \\
         & S3 & 51.6 $\to$ 51.3 & 4.5 $\to$ 4.5 \\
\midrule
\multicolumn{4}{l}{\emph{LLM agents}} \\
Codex    & S1 & 63.2 $\to$ 63.3 & 53.1 $\to$ 52.7 \\
         & S2 & 58.9 $\to$ 58.8 & 50.4 $\to$ 49.7 \\
         & S3 & 61.3 $\to$ 62.0 & 52.0 $\to$ 51.6 \\
Claude Code & S1 & 68.8 $\to$ 68.6 & 21.9 $\to$ 21.6 \\
            & S2 & 65.9 $\to$ 66.6 & 23.6 $\to$ 23.2 \\
            & S3 & 66.2 $\to$ 66.7 & 25.0 $\to$ 24.9 \\
\bottomrule
\end{tabular}
\caption{Performance before and after excluding never-merged
PRs (Full $\to$ Merged-only). All values are percentages.
LLM models are evaluated on 15,895 and 15,096 snapshots,
respectively. For agents, filtering retains 500 positive
and 416 negative examples per stage, except for Codex
Stage 3, which retains 500 positive and 399 negative
examples. The same scoring rules are used before and
after filtering.}
\label{tab:merged-only-results}
\end{table}

\section{Manual Audits}
\label{app:manual-audits}

\subsection{Review Evidence in Never-Merged PRs}

We manually audited 30 randomly sampled never-merged PRs
to examine the technical concerns underlying their historical
outcomes. For each PR, we inspected its review history and
evaluated snapshot to determine whether a technical concern
was documented, whether it applied to the snapshot, and
whether it was supported by the benchmark evidence under
our eight review axes.

We found documented technical concerns in 20 of the 30 PRs.
Among these, 16 concerns applied to the evaluated snapshots,
and 15 were supported by the supplied benchmark evidence.
Cases without documented technical concerns are not
necessarily merge-ready, but their historical outcomes
cannot be attributed to verified snapshot-visible issues.
Complete annotations and supporting GitHub evidence are
available in our
\href{https://huggingface.co/datasets/MathlibPR/MathlibPR/blob/main/rebuttal_audit/human_audits/MathlibPR_manual_audits_30.csv}
{released audit records}.

\subsection{Alignment with Reviewer Feedback}

We also audited 12 correct Stage-3
\texttt{not\_merge\_ready} predictions with substantive
rationales, selecting three each from Codex, Claude Code,
DeepSeek, and Goedel. We compared the model-generated
rationales with documented reviewer feedback and subsequent
PR revisions.

Codex directly identified the documented concern in two
cases and partially aligned in one. Claude Code partially
aligned in all three cases, while the six DeepSeek and
Goedel rationales were unrelated to or unsupported by
the documented concerns. These results illustrate that a
correct verdict does not necessarily imply correct review
reasoning. The audit is qualitative and does not estimate
overall rationale accuracy. Case-level comparisons and
supporting evidence are available in our
\href{https://huggingface.co/datasets/MathlibPR/MathlibPR/blob/main/rebuttal_audit/human_audits/MathlibPR_rationale_audits_12.csv}
{released rationale audit}.

\end{document}